# ReadsClean: a new approach to error correction of sequencing reads based on alignments clustering.

Oleg Fokin[1], Anastasia Bakulina[1,2], Igor Seledtsov[1], and Victor Solovyev[1]

[1]Softberry Inc., Mount Kisco, NY, USA; [2]Novosibirsk State University, Novosibirsk, Russia

**Motivation:** Next generation methods of DNA sequencing produce relatively high rate of reading errors, which interfere with *de novo* genome assembly of newly sequenced organisms and particularly affect the quality of SNP detection important for diagnostics of many hereditary diseases. There exists a number of programs developed for correcting errors in NGS reads. Such programs utilize various approaches and are optimized for different specific tasks, but all of them are far from being able to correct all errors, especially in sequencing reads that crossing by repeats and DNA from di/polyploid eukaryotic genomes.

**Results:** This paper describes a novel method of error correction based on clustering of alignments of similar reads. This method is implemented in ReadsClean program, which is designed for cleaning Illumina HiSeq sequencing reads. We compared ReadsClean to other reads cleaning programs recognized to be the best by several publications. Our sequence assembly tests using actual and simulated sequencing reads show superior results achieved by ReadsClean.

**Availability and implementation:** ReadsClean is implemented as a standalone C code. It is incorporated in an error correction pipeline and is freely available to academic users at Softberry web server www.softberry.com.

**Supplementary information:** (included)..

## Introduction

Next generation sequencing methods have revolutionized the field of genomic research, dramatically shortening time and decreasing cost of sequencing compared to previous-generation methods. Resequencing of genomes, including human genome, and *de novo* sequencing of bacterial genomes have become routine tasks. Still, *de novo* assembly of eukaryotic genomes, much larger and more complex than bacterial ones, remains a difficult task and requires a lot of bioinformatics work. It is especially difficult to correctly assemble sequences of regions containing repeats that differ by a few nucleotides per copy: such differences are routinely subjected to erroneous correction. As repeats comprise up to 40% of eukaryotic genomes, it is very important to improve performance of genome assembling algorithms on such sequences. One of the ways to facilitate this process is to correct errors in sequencing reads before their assembly into contigs. It was previously shown that employing the programs of error correction visibly improves quality of contig assembly. Specifically, value of N50 (meaning that 50% of genome is covered by contigs of at least that length) increases substantially (Salzberg *et al.*, 2012).

Different sequencing platforms generate different types of errors, and this has to be taken into account when choosing an error correction program. This paper covers Illumina HiSeq platform. For a number of years, Illumina remains a leading manufacturer of sequencing machines, and its HiSeq series is currently the most common sequencing platform in the world. It is designed for sequencing entire large genomes: for instance, one run of HiSeq 4000 yields up to five billion reads. Illumina HiSeq reads are usually 100-150 bp long, are paired, and have quality of ≥75% bp

> Q30, i.e. for at least 75% of bases have an error rate of no more than 0.1% (Illumina, 2015). Overwhelming majority of that platform's errors are single nucleotide substitutions.

Despite relatively frequent occurrence of errors in reads, redundant coverage of a genome by reads usually provides enough information to correct most of them. The majority of algorithms of error correction can be divided into the following three groups:

    1. k-spectrum-based. These methods are based on building a spectrum of k-mers (sequences of k nucleotides in a read). Rarely found k-mers are assumed to contain an error. Examples of such programs are Musket (Liu *et al.*, 2013), BLESS (Heo *et al.*, 2014) and Blue (Greenfield *et al.*, 2014).

    2. Suffix tree/array-based. A suffix tree is built, and its analysis finds rare, and therefore putatively erroneous, variants. Examples are RACER (Ilie and Molnar, 2013), SHREC (Schröder *et al.*, 2009) and HiTEC (Ilie *et al.*, 2011).

    3. Multiple sequence alignment-based. These methods utilize various methods of aligning the reads, and then correct mismatches in these alignments. Examples are Coral (Salmela and Schröder, 2011) and Karect (Allam *et al.*, 2015).

Recently developed Karect program (Allam *et al.*, 2015) treats each read as a reference, performs multiple alignment for a set of reads similar to a reference read, stores accumulated results in a partial order graph, computes weights for the graph edges, and construct corrected reads. Karect consistently outperforms other existing techniques in terms of both individual-bases error correction (up to 10% accuracy gain) and post-*de novo* assembly quality (up to 10% increase of NGA50). To reduce computational cost during an alignment phase, the algorithm selects a subset (~ 30) of possible candidate reads and uses them to generate reference read corrections. Such procedure often selects a set of reads from different clusters of repeated sequences, which comprise a very significant fraction of eukaryotic genomes. As a consequence of such selection, a reference read can be corrected by reads from a wrong family of repeats. To overcome this problem, we developed a new error correction algorithm that initially clusterizes overlapping reads and then corrects a reference read using the closest cluster. The algorithm is implemented in ReadsClean program, which can be classified as multiple sequence alignment-based. When used on test sets of NGS reads, ReadsCean outperforms existing techniques in terms of both individual-base error correction and *de novo* assembly of corrected reads into contigs.

## *Methods*

### **Description of Algorithm**

ReadsClean is able to simultaneously process several sets of reads. In a configuration file, user specifies type of each set: mate pair, paired-end, single, maximum length and a distance between paired reads. It is possible to use a set of reads to clean another set, which is useful when coverage of a genome by a set to be cleaned is relatively low, so such set cannot effectively be used to clean itself. Another potential example is a set that contains unusually large number of errors: Then it may make sense to use reads of a cleaner set (cleaning database) for alignment with reads that we intend to correct. Using a specific set for cleaning also allows splitting a set into several subsets and processing them in parallel on different computes (nodes), while using an entire initial set as a cleaning database. All these situations are routinely encountered while working with sequencing reads of large eukaryotic genomes.

The error-correction (cleaning) algorithm consists of the following principal steps (Fig. 1):

1. Creating a multiple alignment of a read being cleaned.
2. Frequency-based correction of obvious errors.
3. Clustering reads in a multiple alignment.
4. Correcting random errors in clusters and possible second clustering.
5. Choosing a cluster for a read to use in correction.
6. Recording the results.

**Figure 1.** Diagram of work flow to correct errors in reads (cleaning reads) by ReadsClean program.

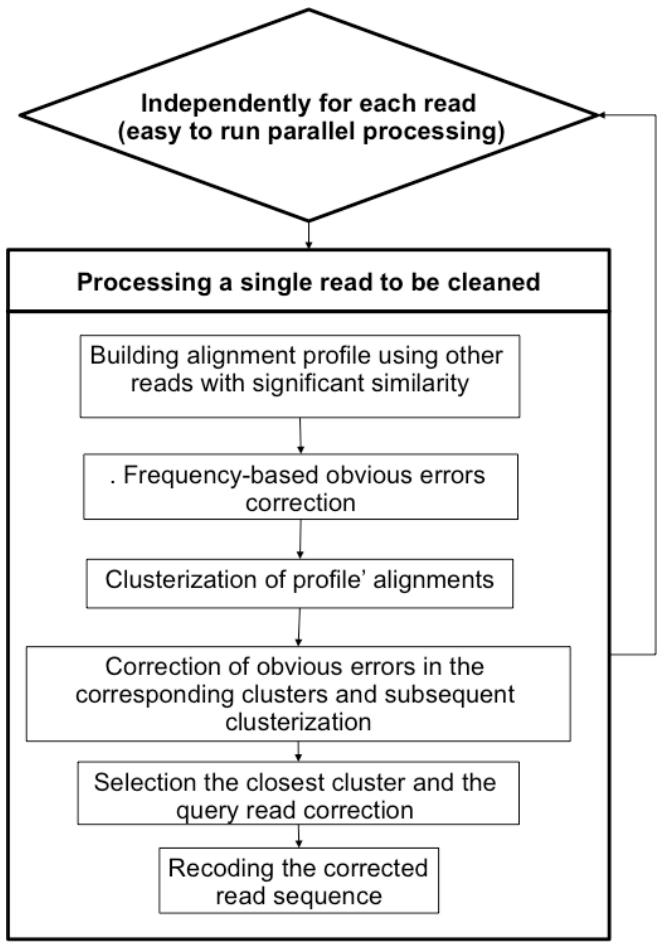

## Building Multiple Alignment for a Read Being Cleaned

Reads from a reference set are aligned with a read to be cleaned (query read). We consider only alignments without internal gaps. To expedite selection of reads for alignment, the reads are hashed in advance. If each pairwise alignment satisfies preset criteria of similarity, such alignment is added to a multiple alignment. Regions of aligned sequences not overlapping with a query read are discarded (Fig. 2A).

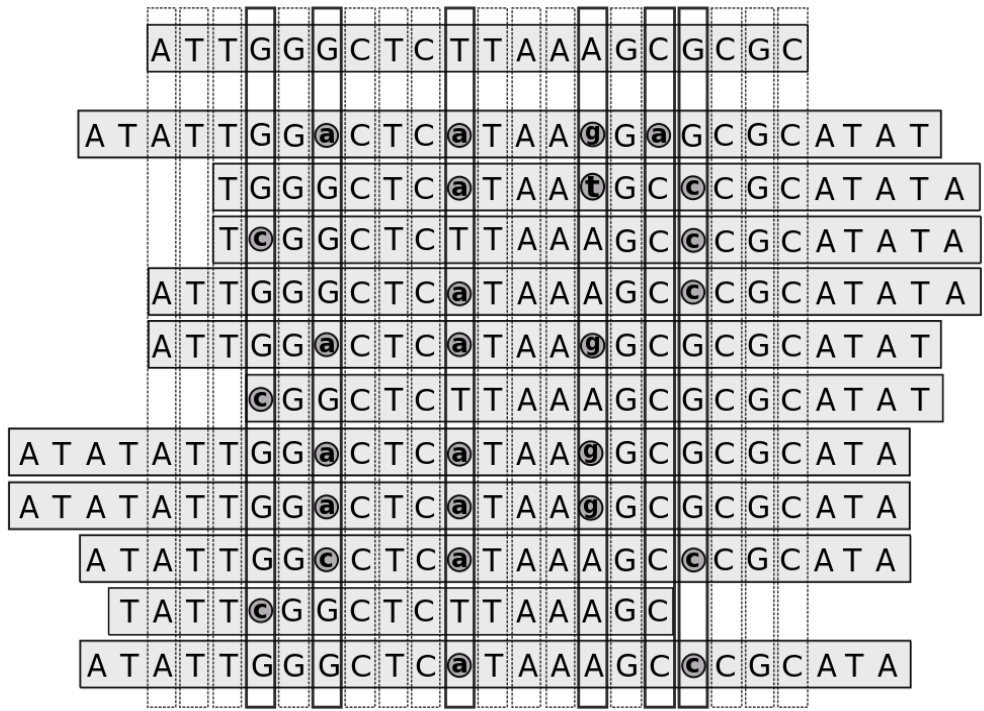

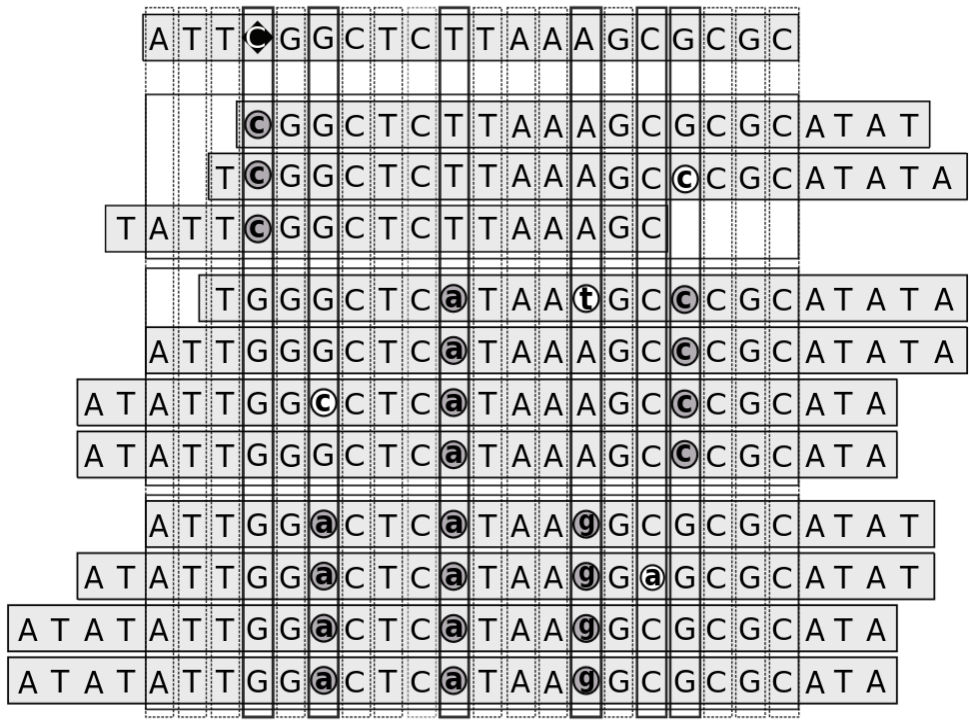

**Figure 2**. A. An example of multiple alignment. Upper line is a query read (read being cleaned). Mismatches with that read are shown in grey circles, small letters. Positions (columns in the alignment) where mismatches are found are shown in boxes. B. The same alignment after clustering and error correction. There are three clusters shown. Mismatches that are excluded

from the subsequent clustering are highlighted in clear circles. A cluster on top, exhibiting the highest similarity to a query read, is used for correcting errors in the read. Black rhombus shows a corrected error. Read tails that do not align to the query read are ignored.

**Frequency-Based Correction of obvious errors**

At this stage, each column of a multiple alignment is analyzed. If, in a given position, the most frequent nucleotide is found at least N times, and each of remaining nucleotides no more than M times, all less frequent nucleotides are replaced by the most frequent. By default, N=7 and M=1. Such positions are marked to be excluded from subsequent clustering.

**Clustering of Reads in Multiple Alignment**

Reads in an alignment are clustered using descending hierarchical clustering method. A single position is chosen where distribution of frequencies of nucleotides is closest to random, reads in the alignment are sorted into clusters corresponding to a common nucleotide found in that position, and the position is excluded from the subsequent clustering. In each cluster, positions that are dominated by a single nucleotide (determined by the same way as in frequency-based error correction) are also excluded from the subsequent clustering. Then each cluster is split into subclusters based on the same rules and so forth. The process continues until all positions in the alignment are marked to be excluded from further clustering.

**Correction of reads in clusters and second clustering**

A consensus sequence is determined for each cluster, and query reads are corrected in order to correspond to the selected consensus. Positions where the consensus nucleotide is found only a few times (by default, less than five times) are marked as unreliable. Then the corrected reads are subjected to a second round of clustering. In many cases, clusters remain unchanged, but second clustering somewhat improves accuracy in our tests, while not requiring too much time to complete (data not shown).

**Choosing cluster for read correction**

The alignment clusters are sorted by similarity of their consensus sequences to a query read, and also by their lengths, from longest to shortest. Clusters covering different regions of a query read can be merged if they overlap by at least 30%, and if their consensus sequences do not contradict each other. For such merged clusters, a new consensus sequence is determined. For correcting a read, a cluster with highest similarity of its consensus sequence, which does not contain unreliable positions, is chosen. If there are several clusters of same similarity, one that contains higher number of reads is chosen (Fig. 2B).

**Recording the results**

If, for a given query read, no similar clusters are found covering its entire length and without unreliable positions, or if there are several clusters with roughly the same similarity, such read is still corrected, but it is marked as unreliable. The program has an option of recording unreliable reads into a separate file. If one of paired reads is marked unreliable, its reliable pair is recorded in a file with single reads. If an alignment can't be built for a query read, or if no acceptably similar clusters were found, such read is not corrected and also recorded in a file of unreliable reads. Theoretically, non-reliable reads can be re-cleaned by using newly cleaned reliable reads as a reference. Our tests, however, show that this doesn't improve the results.

For *de novo* genome assembly, it is sometimes desirable to use all reads, including unreliable ones, especially if they constitute more than 1% of all reads, because discarding unreliable reads would reduce sequencing coverage. But for SNP finding, it is better to avoid unreliable reads, in order to reduce the number of false positives.

If the number of unreliable reads is less than 1%, we recommend discarding them for genome assembly computing. If the number is between 1% and 5%, both variants of assembly (with and without unreliable reads) can be attempted.

Heterozygosity of eukaryotic genomes doesn't pose a big problem for our approach, as each allele is usually sorted into a separate cluster in a cleaning set, and the algorithm corrects each read individually and independently from other reads.

## Results

### Testing procedure

The following programs were used to compare error correction accuracy: RACER (Ilie and Molnar, 2013), SGA (Simpson and Durbin, 2012), Musket (Liu et al., 2013), Coral (Salmela and Schröder, 2011), Karect (Allam et al., 2015). These programs were acknowledged to be the best in previously published reviews and publications (Allam et al., 2015, Alic et al., 2016). The parameters settings that used to run these programs are listed in the Supplement.

To compare error correction programs, we used both simulated and real sequencing data. Simulated data were obtained with SimSeq program (https://github.com/jstjohn/SimSeq). That program introduces errors specific to Illumina platform, with the average number of errors around 0.9%. We generated simulated reads for E.coli (1,855,870 PE reads, 100 bp read length, 300 bp distance between paired reads, 0.936% errors) and human chr14 (35,315,412 PE reads, 100 bp length, 300 bp distance between paired reads, 0.962% errors). These sets represented about 40x genome coverage. Using our own software, we also introduced additional errors into chr14 simulated reads, to obtain a set with error frequencies up to 3%. The results of reads cleaning provide information on the number of corrected errors, missed errors, errors replaced by other errors, and introduced errors.

As the real sequenced reads, we used data from GAGE project (Salzberg et al., 2012) for the following genomes: Staphylococcus aureus, Rhodobacter sphaeroides, and human chromosome 14. In all cases, only files from Library 1 were used. They contain paired reads in fastq format with the following parameters. S. aureus: average read length: 101 bp, insert length: 180 bp, number of reads: 1,294,104. R. sphaeroides: average read length: 101 bp, insert length: 180 bp, number of reads: 2,050,868. H. sapiens Chr 14: average read length: 101 bp, insert length: 155bp, number of reads: 36,504,800. To evaluate performance of read correction software, *de novo* sequence assembly was performed, as its results give some indication of usefulness of error correction methods. The programs used for assembly were SGA (Simpson and Durbin, 2012),

Velvet (Zerbino and Birney, 2008), SOAPdenovo (Luo et al., 2012) and SPADES (Bankevich et al., 2012). Parameters settings of these programs are listed in the Supplement. Quality of assembly was estimated using QUAST program (Gurevich et al., 2013).

**Correction quality evaluation on simulated reads**

The results of tests on simulated data are shown in Tables 1-3. Since Coral program not only corrects errors, but also discards reads deemed low quality, its results are not presented. The table shows the corrected errors, errors left after correction, missed errors, errors replaced by other errors, and introduced errors.

**Table 1**. Simulated *E.coli* reads, initially containing 1,737,154 (0.936%) errors. Absolute number and percentages (of initial number of errors) are shown.

| Program | Corrected errors | Errors left after correction | Missed errors | Errors replaced by errors | Introduced errors |
|---|---|---|---|---|---|
| ReadsClean | | | | | |
| | **1,736,962** | **265** | **147** | 45 | **73** |
| | **99.9889%** | **0.0153%** | **0.0085%** | 0.0026% | **0.0042%** |
| RACER | 1,734,832 | 22,644 | 2,175 | 147 | 20,322 |
| | 99.87% | 1.30% | 0.13% | 0.01% | 1.17% |
| SGA | 1,733,562 | 4,550 | 3,552 | **40** | 958 |
| | 99.79% | 0.26% | 0.20% | **0.002** | 0.06% |
| Musket | 1,731,964 | 5,620 | 5,013 | 177 | 430 |
| | 99.70% | 0.32% | 0.29% | 0.01% | 0.02% |
| Karect | 1,736,858 | 784 | 217 | 79 | 488 |
| | 99.98% | 0.05% | 0.01 | 0.005% | 0.03% |

**Table 2**. Simulated human chromosome 14 reads, initially containing 33,977,454 (0.962%) errors. Absolute numbers and percentages (of initial number of errors) are shown.

| Program | Corrected errors | Errors left after correction | Missed errors | Errors replaced by errors | Introduced errors |
|---|---|---|---|---|---|
| ReadsClean | | | | | |
| | **33,909,951** | **119,617** | **62,434** | **5,069** | **52,114** |
| | **99.8013%** | **0.3520%** | **0.1838%** | **0.0149%** | **0.1534%** |
| RACER | 31,242,161 | 22,157,432 | 2,588,218 | 147,075 | 19,422,139 |
| | 91.95% | 65.21% | 7.62% | 0.43% | 57.16% |
| SGA | 31,511,057 | 3,294,087 | 2,436,166 | 30,231 | 827,690 |
| | 92.74% | 9.69% | 7.17% | 0.09% | 2.44% |
| Musket | 30,533,489 | 4,151,370 | 3,354,877 | 89,088 | 707,405 |
| | 89.86% | 12.22% | 9.87% | 0.26% | 2.08% |
| Karect | 32,573,973 | 1,676,124 | 1,359,854 | 43,627 | 272,643 |
| | 95.87% | 9.87% | 4.00% | 0.13% | 0.80% |

**Table 3.**  Simulated human chromosome 14 reads, initially containing 106,213,917 (about 3%) errros.  Absolute numbers and percentages (of initial number of errors) are shown.

| Program | Corrected errors | Errors left after correction | Missed errors | Errors replaced by errors | Introduced errors |
|---|---|---|---|---|---|
| ReadsClean | | | | | 2,254,036 |
| | | | | | 2.1222% |
| | **105,758,530** | **2,287,921** | **332,354** | **123,033** | 1,832,534 |
| | **99.5713%** | **2.1541%** | **0.3129%** | **0.1158%** | 1.7253% |
| RACER | 97,200,336 | 51,332,436 | 6,982,319 | 2,031,262 | 42,318,855 |
| | 91.51% | 48.33% | 6.57% | 1.91% | 39.84% |
| SGA | 93,661,600 | 14,632,878 | 10,709,063 | 1,843,254 | 2,080,561 |
| | 88.18% | 13.78% | 10.08% | 1.74% | 1.96% |
| Musket | 90,551,567 | 18,532,282 | 15,365,822 | 296,528 | 2,869,932 |
| | 85.25% | 14.47% | 14.47% | 0.28% | 2.70% |
| Karect | 101,477,152 | 5,534,059 | 4,490,772 | 245,993 | **797,294** |
| | 95.54% | 5.21% | 4.23% | 0.23% | **0.75%** |

The results in Tables 1-3 demonstrate that all programs correct the majority of errors. It can be seen, however, that the ReadsClean corrects more errors on human chr14 reads than other programs, while also leaving fewer miscorrected and introduced errors. ReadsClean also performs better then the other programs on human chr14 reads with 3% of errors. On *E.coli* data, ReadsClean and Karect miss comparable percentage of errors, but Karect introduces more new errors, while other programs perform markedly worse than these two. In short, ReadsClean shows superior performance on all three sets of simulated reads.

### Applying error correction to improve sequence assembling

We compared contig assemblies from uncorrected reads and those cleaned by different error correction programs. Many *de novo* assembly programs have their own built in algorithms of cleaning and filtering reads, for instance MaSuRCA (Zimin *et al*. 2013). We used the programs that lack such option, or where it can be switched off. For SGA and Velvet programs, we used the same settings as GAGE project. For SOAPdenovo, its parameters settings were adopted from the program manual, as we used newer version than GAGE project. For SPADES (Bankevich *et al*., 2012), we used default settings and turned off the error correction step.

As criteria of assembly quality, we used the following measures: NGA50 (similar to N50, but takes into account sizes of alignments of contigs to genome rather than simply sizes of contigs, thus excluding from consideration misassembled contigs), genome coverage – a fraction of genome covered by contigs. Human chr14 contigs were aligned to the chr14 sequence, which contains many unknown positions (N in a sequence), therefore actual genome coverage cannot exceed 82.24%. We excluded these N-regions from consideration in these tests, that allowing coverage to go beyond that limit.

**Table 4.** Results of assembling simulated reads of *E.coli*.

| Criterion | Cleaning Program | SGA | Velvet | SOAPdenovo | SPADES |
|---|---|---:|---:|---:|---:|
| Percent of genome coverage | No cleaning | **97.887** | 97.863 | 97.938 | 98.310 |
| | ReadsClean | 97.883 | **97.938** | **97.979** | 98.307 |
| | RACER | 97.783 | 97.833 | 97.960 | 98.306 |
| | SGA | 97.834 | 97.876 | 97.957 | 98.306 |
| | Musket | 97.844 | 97.837 | 97.964 | 98.307 |
| | Coral | 97.860 | 97.786 | 97.808 | 98.054 |
| | Karect | 97.864 | 97.845 | 97.901 | **98.323** |
| NGA50 | No cleaning | 29,152 | **107,662** | 46,442 | **133,026** |
| | ReadsClean | **54,946** | 107,638 | **82,580** | **133,026** |
| | RACER | 31,662 | 107,638 | **82,580** | **133,026** |
| | SGA | 42,717 | 107,637 | **82,580** | **133,026** |
| | Musket | 53,643 | 107,638 | **82,580** | **133,026** |
| | Coral | 52,578 | 105,532 | 65,530 | 123,773 |
| | Karect | 54,943 | 107,638 | **82,580** | **133,026** |

For E.coli bacterial reads (Table 4) it can be seen that cleaning reads before assembly doesn't yield obvious benefits for all assembling programs. For Velvet and SPADES programs, the results are unchanged, while for SGA and SOAPdenovo, NGA50 increases about the same no matter which cleaning program was used.

**Table 5**. Results of assembling simulated reads of human chr14 with 0.962% of errors.

| Criterion | Cleaning Program | SGA | Velvet | SOAPdenovo | SPADES |
|---|---|---:|---:|---:|---:|
| Percent of genome coverage | No cleaning | **90.328** | 84.643 | 95.813 | 96.726 |
| | ReadsClean | 89.322 | **88.058** | **96.086** | **96.833** |
| | RACER | 83.783 | 84.449 | 95.323 | 96.605 |
| | SGA | 86.800 | 85.085 | 95.798 | 96.733 |
| | Musket | 88.814 | 84.585 | 95.640 | 96.750 |
| | Coral | 87.796 | 85.909 | 92.369 | 93.824 |
| | Karect | 88.912 | 86.101 | 95.460 | 96.747 |
| NGA50 | No cleaning | **4,068** | 4,808 | 10,438 | 19,415 |
| | ReadsClean | 4,057 | **10,553** | **13,069** | **20,418** |
| | RACER | 2,698 | 4,364 | 10,027 | 17,299 |
| | SGA | 3,284 | 5,449 | 11,821 | 18,577 |
| | Musket | 3,931 | 4,731 | 11,004 | 19,654 |
| | Coral | 3,823 | 7,549 | 6,587 | 9,174 |
| | Karect | 3,987 | 6,671 | 10,061 | 18,385 |

**Table 6**. Results of assembling simulated reads of human chr14 with 3% of errors.

| Criterion | Cleaning Program | SGA | Velvet | SOAPdenovo | SPADES |
|---|---|---|---|---|---|
| Percent of genome coverage | No cleaning | 0.758 | 5.331 | 9.015 | 93.909 |
| | ReadsClean | **89.323** | **87.686** | **95.846** | **96.745** |
| | RACER | 78.899 | 82.068 | 93.959 | 94.850 |
| | SGA | 85.953 | 82.574 | 94.434 | 95.164 |
| | Musket | 86.939 | 81.980 | 93.843 | 95.025 |
| | Coral | 87.702 | 85.547 | 92.706 | 94.244 |
| | Karect | 88.763 | 84.160 | 94.902 | 95.832 |
| NGA50 | No cleaning | - | - | - | 7,418 |
| | ReadsClean | **4,058** | **10,203** | **11,693** | **18,812** |
| | RACER | 2,050 | 3,022 | 6,502 | 8,987 |
| | SGA | 3,104 | 3,168 | 7,162 | 9,468 |
| | Musket | 3,898 | 3,075 | 6,559 | 9,504 |
| | Coral | 3,770 | 7,235 | 6,736 | 10,148 |
| | Karect | 3,940 | 3,922 | 8,690 | 11,911 |

On human data (Table 5) ReadsClean performs the best with all assembly programs except SGA, which has by far the lowest value of NGA50 of all four assembling programs. These data clearly show assembly improvement due to error correction. As the number of errors increase (Table 6), all assembly programs except SPADES perform very poorly on uncleaned reads, and all programs show significant improvement on cleaned reads. In this test, ReadsClean outperformed the other cleaning programs with all assemblers, in both sequence coverage and NGA50.

## Tests on a set of real data generated by sequencing machine

For real data, the actual percentage of errors is not known. For that reason, the quality of error correction programs was compared only by the results of *de novo* assembly. We used the same assembly programs and same criteria as for simulated data. The results of tests using actual reads from GAGE project are shown in Tables 7-9.

**Table 7**. Results of *de novo* assembly of reads of *Staphylococcus aureus* genome.

| Criterion | Cleaning Program | SGA | Velvet | SOAPdenovo | SPADES |
|---|---|---|---|---|---|
| Percent of genome coverage | No cleaning | 81.540 | 96.311 | 84.741 | **98.489** |
| | ReadsClean | 97.190 | **97.001** | **97.767** | 98.443 |
| | RACER | 93.537 | 95.940 | 93.820 | 98.034 |
| | SGA | 94.402 | 96.672 | 95.571 | 98.335 |
| | Musket | 96.308 | 96.567 | 97.283 | 98.381 |
| | Coral | 96.099 | 96.696 | 96.322 | 98.029 |
| | Karect | **97.314** | 96.989 | 97.760 | 98.455 |
| NGA50 | No cleaning | 1,263 | 14,984 | 1,369 | 46,907 |
| | ReadsClean | 13,022 | 36,667 | 8,977 | 55,044 |
| | RACER | 3,553 | 16,835 | 2,901 | 40,860 |
| | SGA | 4,005 | 23,156 | 3,976 | 53,695 |
| | Musket | 7,481 | 23,155 | 7,050 | 55,906 |
| | Coral | 7,125 | 29,840 | 5,931 | 46,527 |
| | Karect | **20,250** | **40,816** | **13,223** | **69,745** |

**Table 8.** Results of *de novo* assembly of reads of *Rhodobacter sphaeroides* genome.

| Criterion | Cleaning Program | SGA | Velvet | SOAPdenovo | SPADES |
|---|---|---:|---:|---:|---:|
| Percent of genome coverage | No cleaning | 47.196 | 93.556 | 65.762 | 98.514 |
| | ReadsClean | 86.077 | 93.745 | 88.451 | 98.528 |
| | RACER | 84.379 | 87.299 | 86.298 | 97.922 |
| | SGA | 87.914 | 93.165 | 90.304 | **98.625** |
| | Musket | 87.896 | 93.887 | 88.431 | 98.167 |
| | Coral | 83.870 | 92.958 | 84.647 | 98.075 |
| | Karect | **94.183** | **95.777** | **94.333** | 98.555 |
| NGA50 | No cleaning | - | 5,994 | 724 | 15,344 |
| | ReadsClean | 1,977 | 7,424 | 1,700 | 15,284 |
| | RACER | 1,676 | 2,010 | 1,625 | 9,054 |
| | SGA | 2,202 | 6,834 | 2,040 | 17,542 |
| | Musket | 1,996 | 3,462 | 1,831 | 12,756 |
| | Coral | 1,679 | 6,728 | 1,430 | 15,099 |
| | Karect | **3,855** | **10,558** | **3,037** | **21,044** |

**Table 9.** Results of *de novo* assembly of reads of human chromosome 14.

| Criterion | Cleaning Program | SGA | Velvet | SOAPdenovo | SPADES |
|---|---|---:|---:|---:|---:|
| Percent of genome coverage | No cleaning | 70.328 | 65.937 | 75.443 | 78.512 |
| | ReadsClean | **71418** | **67.956** | **76.318** | **79.154** |
| | RACER | 65.873 | 65.372 | 74.768 | 77.758 |
| | SGA | 69.679 | 66.637 | 76.079 | 78.889 |
| | Musket | 70.061 | 65.365 | 75.741 | 78.727 |
| | Coral | 69.739 | 66.723 | 73.467 | 76.224 |
| | Karect | 70.992 | 67.527 | 75.598 | 79.020 |
| NGA50 | No cleaning | 2,028 | 1,844 | 2,604 | 5,401 |
| | ReadsClean | **2,283** | **2,247** | **3,204** | **10,384** |
| | RACER | 1,434 | 1,633 | 2,490 | 5,211 |
| | SGA | 1,971 | 1,911 | 3,118 | 9,067 |
| | Musket | 2,064 | 1,616 | 2,892 | 8,393 |
| | Coral | 2,082 | 1,984 | 2,545 | 5,293 |
| | Karect | 2,271 | 2,114 | 2,980 | 10,048 |

Tables 7-9 show that ReadsClean outperforms the other tools on data from human chromosome 14, while for smaller bacterial genomes Karect and RedsClean lead in assembly quality.

## *Conclusion*

The modern sequencing technologies always generate sequencing reads with some percentage of errors, so subsequent error correction is becoming a routine stage of handling NGS data. Our tests demonstrate that reads cleaning is less important for assembling short bacterial genomes with good coverage, however for eukaryotic sequencing data, especially with high level of errors, it significantly improves the results of genome assembling. ReadsClean outperforms its peers on such datasets and can be recommended for projects involving *de novo* assembling eukaryotic genomes, and it can be used for error correction in polyploid genomes where its is very essential to differentiate SNPs from sequencing errors. Also, our clustering approach is very useful for error correction in genomes containing multiple groups of repeated sequences, when the correction must be done within the corresponding repeat cluster. ReadsClean algorithm is easy to run in parallel mode, as each read is corrected independently from others, and a version exists that is optimized for very fast performance on an Amazon Web Services (AWS) cluster.

In the future, it is possible to adopt the program for cleaning reads generated by platforms other than Illumina, which will require additional tests and adjustments of settings. The program is freely available to download for academic users at the Softberry web server: http://www.softberry.com/berry.phtml?topic=fdp.htm&no_menu=on

**Supplementary information:**

**Parameters for running error-correction programs to test their accuracy for Human chr 14 reads from GAGE project:**

PE reads located in two files: frag_1.fastq и frag_2.fastq, distance between paired reads 155 bp, approximate size of chromosome 100000000 bp.

**readsClean**
```
./sb.pl -cfg:config.cfg -dir:mydir -clean -j:4
```

config.cfg:

```
pe.pe1 {
  --cleaning
  src.file = ["frag_1.fastq"  "frag_2.fastq"]
  cleaning {
    --addition-clean
    --single-out
  }
  dst.format = fasta
  dst.count  = 1
}
```

**RACER**
```
../RACER frag_1.fastq output_1.fastq 100000000
../RACER frag_2.fastq output_2.fastq 100000000
```

**SGA**
```
sga preprocess  -p 1 frag_1.fa frag_2.fa > frag.pp.fa
sga index -t 20 frag.pp.fa
```

```
sga correct -k 31 -t 20 frag.pp.fa -o output.fa
```

**Musket**
```
musket frag_1.fastq -o output_1.fastq -inorder
musket frag_2.fastq -o output_2.fastq -inorder
```

**Coral**
```
coral -fq frag_1.fastq -o output_1.fastq -illumina
coral -fq frag_2.fastq -o output_2.fastq -illumina
```

**Karect**
```
./karect -correct -inputfile=frag_1.fastq -inputfile=frag_2.fastq  -matchtype=hamming -celltype=diploid
```

**Parameters for running sequence assembling programs:**

PE reads located in two files: output_1.fastq output_2.fastq, distance between paired reads 155 bp:

**SGA**
```
/sf/gen7/gen5/SNP/cleaner/SGA/bin/sga preprocess -p 1 output_1.fastq outout_2.fastq > frag.pp.fa
/sf/gen7/gen5/SNP/cleaner/SGA/bin/sga index -t 31 frag.pp.fa
/sf/gen7/gen5/SNP/cleaner/SGA/bin/sga filter frag.pp.fa
/sf/gen7/gen5/SNP/cleaner/SGA/bin/sga overlap -t 20 -m 45 frag.pp.filter.pass.fa
/sf/gen7/gen5/SNP/cleaner/SGA/bin/sga assemble frag.pp.filter.pass.asqg.gz
```

**Velvet**
```
velveth . 31 -fastq -shortPaired -separate output1.fastq output2.fastq
velvetg . -exp_cov auto -ins_length 155 -ins_length_sd 15
```

**SOAPdenovo**
```
SOAPdenovo-63mer all -s soap.config -K 63 -R -o soap -p 30
```

soap.config:

```
# maximal read length
max_rd_len=101

[LIB]
# average insert size
avg_ins=155

# if sequence needs to be reversed
reverse_seq=0

# in which part(s) the reads are used
asm_flags=1

# use only first 50 bps of each read
rd_len_cutoff=101
```

```
# in which order the reads are used while scaffolding
rank=1

# cutoff of pair number for a reliable connection (default 3)
pair_num_cutoff=3

# minimum aligned length to contigs for a reliable read location (default 32)
map_len=32

# fastq file for read 1
q1=output_1.fastq

# fastq file for read 2 always follows fastq file for read 1
q2=output_2.fastq
```

## SPADES
```
spades.py -1 output_1.fastq -2 output_2.fastq --only-assembler -o spades
```